
\magnification=1200
\baselineskip=24pt plus .5pt minus .5pt

{\it Submitted to Physical Review E 6/30/1993}
\centerline{\bf Grain Boundary Buckling and Spin Glass Models of
Disorder in Membranes}
\bigskip
\centerline{Carlo Carraro and David R. Nelson}
\medskip
\centerline
{\it Department of Physics}
\centerline {\it Harvard University}
\centerline
{\it Cambridge, MA 02138}
\medskip
\bigskip

\medskip
\bigskip
{\narrower\narrower\smallskip
A systematic investigation is presented of grain boundaries and
grain boundary networks in two dimensional flexible membranes with crystalline
order.  An isolated grain boundary undergoes a buckling
transition at a critical value of the elastic constants, but, contrary to
previous findings, the shape of the buckled membrane is shown to be
asymptotically flat. This is in general not true
in the case of a network of intersecting grain boundaries, where each
intersection behaves as the source of a long range stress field.
Unlike the case of isolated dislocations or
disclinations, the energy associated with these stresses is
finite; they can, however, destabilize the flat phase. The
buckled phase can be modeled by an Ising spin-glass with long range
antiferromagnetic interactions.
These findings may be relevant to the understanding of the wrinkling transition
in partially polymerized vesicles reported in recent experiments.
\smallskip}
\medskip
{PACS numbers: 87.22.Bt, 46.30.Lx}
\goodbreak
\bigskip
\centerline{\bf I. INTRODUCTION}
\medskip
Flexible membranes are two dimensional (2D) generalizations of
linear polymer chains. The properties of a 2D membrane, embedded in three
dimensional space, depend strongly on the internal order, crystalline,
hexatic, or fluid. As in other realizations of 2D matter, defects, and their
interactions, affect crucially the stability of a given phase. For example,
the melting transition in two dimensions can be
described as proliferation of topological defects [1].
A membrane, however, is not confined to a plane.
Thus, although the stable phase of a defect-free crystalline
membrane at low temperature
is flat, strains induced by a defect, such as a dislocation, can be
accommodated by displacements in the normal direction, resulting in the
buckling of the membrane [2]. This process entails a trade-off of in-plane
stretching energy for curvature energy, and, hence, it occurs when
$$
{K_0l^2\over\kappa}\ge\gamma.
\eqno(1)
$$
Here, $K_0$ is Young's modulus, $\kappa$ is the bending rigidity, $l$ is a
length scale, and $\gamma$ is a dimensionless constant of order $10^2$.
In membranes of size $R$, $l=R$ for disclinations, and $l=\sqrt{Rb}$ for a
dislocation with Burger's vector $\vec b$.
Thus, these defects always buckle in sufficiently
large membranes, irrespective of the value of the elastic constants.

This conclusion, however, does not hold for finite energy
defects such as vacancies, interstitials, or
tightly bound dislocation pairs. In this case, $l$ is of order a
lattice constant, and thus, the stability of the flat phase in the presence of
such defects is determined by the actual value of the elastic constants. This
brings about the interesting possibility of a buckling transition in an
infinite system as a function of temperature.
Because $\kappa$ typically increases with temperature, while $K_0$
decreases, the condition (1) may be associated with a boundary in the
$(l^2,T)$-plane, separating buckled and flat regimes (see Fig.~1).

Recent studies of membranes with defects and quenched random disorder have
been prompted in part by experiments of Mutz {\it et al.} [3], showing that
partially polymerized vesicles undergo a (possibly first-order) reversible
phase transition from a high temperature phase characterized by a smooth,
``floppy'' surface, to a low temperature phase where the vesicles
appear rigid and
highly wrinkled. This wrinkling transition has been likened to the spin-glass
transition of magnetic systems [3,4].
To understand the mechanism of the transition,
it is helpful to observe that the effect of the partial polymerization
is to nucleate microcrystalline domains around polymerized patches.  Upon
cooling, crystalline order arises in the unpolymerized material surrounding
these domains. This has led to speculation that the grain boundaries,
resulting when microcrystals with different orientations meet, may be
responsible for the observed wrinkled structure. The following argument was
given to support this scenario [5].

Consider a low angle grain boundary in a 2D crystal. This can
be viewed as a row of dislocations, with Burger's vector $\vec b$
perpendicular to the boundary. Let $\theta$ be the boundary
tilt angle. The spacing
between dislocations, $h$, is then given by Frank's law [6],
$h=b/2\sin(\theta/2)\approx b/\theta$.
On length scales large compared to $h$, the stress fields of the dislocations
are screened exponentially, provided the average Burger's vector is
strictly perpendicular; on
shorter length scales, however, the behavior of the grain boundary should be
dominated by the individual dislocations.
Computer simulations of tethered membranes [2] have shown that isolated
dislocations in a finite membrane of size $R$ destabilize the flat phase,
provided that $R>127\kappa/K_0b\equiv R_B$, consistent with Eq.~(1) with
$l=\sqrt{Rb}$.
Hence, one might expect that, if $h>R_B$, the grain boundary will buckle
out of flat space, possibly taking the shape of a rippled hinge, possibly
with large out-of-plane displacements, which would then appear as wrinkles.

The present work is devoted to a systematic study of grain boundaries and
grain boundary networks in two dimensional flexible membranes with crystalline
order. While we find that an infinite grain boundary does indeed undergo
an intriguing
second order transition from a flat to a corrugated phase as the ratio of the
bending rigidity to the shear modulus or
tilt angle is varied, we show that the out-of-plane displacement decays
exponentially away from the boundary, i.e., the shape of the membrane remains
asymptotically flat. We also find, however, that this conclusion does not hold
in the case of a network of intersecting grain boundaries, where each
intersection behaves as the source of a long range stress field, e.g., as a
disclination multipole. Unlike the case of isolated dislocations or
disclinations, the energy associated with these stresses is
generally finite; they can, however, cause buckling at a
sufficiently low temperature, determined by the size $l$ of the (localized)
multipole source.

Thus, upon cooling of a polycrystalline membrane, in which a grain
boundary network has formed, one expects a hierarchy of buckling transitions
to take place, with the low-angle grain boundaries puckering first, followed
by the high-angle ones (low-angle grain boundaries correspond to very large
values of $l^2$ in Fig.~1).
Eventually, as the network of buckled grain boundaries begins to
percolate across the membrane, the (point-like)
grain boundary nodes will buckle, causing large wrinkles to develop fully.
We will show that the long range, often
highly anisotropic, interactions in the buckled phase can favor
buckling of a pair of nodes in opposite directions. Buckling of these
nodes can be described by randomly placed Ising-like degrees of freedom, with
antiferromagnetic interaction. Thus,
frustration will play an important role in the conformation of these membranes
in the buckled phase, much as in spin systems with (long range)
random antiferromagnetic exchange [7,8]. Ising spin-glass
behavior seems likely at low temperature, which may provide a plausible
explanation of the wrinkling transition reported in Ref. [3].

The balance of our paper is as follows. We review some results of the theory
of membrane buckling in Section II, and establish a connection with the Landau
theory of second order transitions. In Section III, we study
isolated grain boundaries and grain boundary networks within the continuum
elastic theory of membranes. Our findings are supported by numerical
simulations presented in Section IV, and we conclude in Section V with a
discussion of simplified models based on this work.

\goodbreak
\bigskip
\centerline{\bf II. BUCKLING OF ELASTIC MEMBRANES}
\medskip
We begin by reviewing the continuum theory of elastic membranes [9].
Let the shape
of a membrane be parameterized by a three dimensional displacement vector with
respect to a reference plane $(x,y)$, and let $\vec u$, $f$ be
the in-plane and out-of-plane components respectively.
The energy of the membrane in a given configuration is the sum of a
stretching and a bending term,
$$
\eqalign{ F&=F_s+F_b\cr
F_s&={1\over 2}\int d^2r\biggl(2\mu u_{ij}^2 + \lambda u_{ii}^2 \biggr)\cr
F_b&={1\over 2}\kappa\int d^2r\biggl(\nabla^2 f\biggr)^2.\cr
}
\eqno(2)
$$
Summations over repeated indices are understood.
Here $\kappa$ is the bending rigidity, $\lambda$ and $\mu$ Lam\'e
coefficients, and the strain tensor is given by
$$
u_{ij}={1\over 2}(\partial_i u_j+\partial_j u_i +\partial_i f
\partial_j f).
\eqno(3)
$$
Note the manifest up-down symmetry $(f\to -f)$.
A given configuration of a fluctuating membrane occurs with
probability proportional to $\exp(-F/k_BT)$.

Minimizing the energy with respect to the displacement yields the equations
$$
\eqalign{
\kappa \nabla^4 f&= \sigma_{ij}\partial_i \partial_j f \cr
\partial_i \sigma_{ij}&=0,
}
\eqno(4)
$$
where the stress tensor $\sigma_{ij}$ is related to the (nonlinear)
strain tensor by
$$
\sigma_{ij}=2\mu u_{ij} + \lambda u_{ll}\delta_{ij}.
\eqno(5)
$$
Because $\sigma_{ij}$ is divergenceless and symmetric, it can be expressed
through a single scalar potential, the Airy stress function $\chi$:
$$
\sigma_{xx}={\partial^2 \chi\over \partial y^2};\ \
\sigma_{yy}={\partial^2 \chi\over \partial x^2};\ \
\sigma_{xy}=-{\partial^2 \chi\over \partial x \partial y}.
\eqno(3)
$$

Equations of motion for the energy functional are most conveniently derived
in terms of $f$ and $\chi$.
In Cartesian coordinates, the equations, obtained originally by von K\'arm\'an,
read
$$
\kappa \nabla^4 f=
{\partial^2 \chi \over \partial y^2}{\partial^2 f \over \partial x^2}+
{\partial^2 \chi \over \partial x^2}{\partial^2 f \over \partial y^2}-
2{\partial^2 \chi \over \partial x \partial y}
{\partial^2 f \over \partial x\partial y}
\eqno(7a)
$$
$$
{1\over K_0} \nabla^4 \chi =
-{\partial^2 f \over \partial x^2}{\partial^2 f \over \partial y^2}+
\biggl({\partial^2 f \over \partial x\partial y}\biggr)^2
+S(\vec r).
\eqno(7b)
$$
Here $K_0=4\mu(\mu+\lambda)/(2\mu+\lambda)$ is the two-dimensional Young's
modulus, and the source term, $S(\vec r)$, is the total density of defects.
For an isolated disclination at $\vec r_0$ with charge $s$, one has
$$
S(\vec r)= s\delta(\vec r - \vec r_0),
\eqno(8a)
$$
while for a dislocation with Burger's vector $\vec b$ we have
$$
S(\vec r)= \epsilon_{ij}b_i \partial_j \delta (\vec r - \vec r_0).
\eqno(8b)
$$
An isolated  vacancy, interstitial, or impurity atom [4] is associated with a
source term
$$
S(\vec r)= {\Omega_0\over 2\pi (1-\sigma)}\nabla^2\delta(\vec r - \vec r_0),
\eqno(8c)
$$
where $\sigma=\lambda/(2\mu+\lambda)$ is the 2D Poisson ratio and $\Omega_0$
the area change associated with the defect.
Other defects may be regarded as dislocation or disclination multipoles.

An effective free energy functional for the normal displacement $f$
can be obtained by integrating out the in-plane displacement
field $\vec u$. The result is [10]
$$
{\cal F}= {1\over 2}K_0
\int d^2r\biggl(\nabla^{-2}\biggl(K(\vec r)-S(\vec r)\biggr)\biggr)^2+
{1\over 2}\kappa\int d^2r\biggl(\nabla^2 f\biggr)^2,
\eqno(9)
$$
where $K(\vec r)\equiv ({\partial^2 f/\partial x^2})
({\partial^2 f / \partial y^2})-({\partial^2 f/\partial x\partial y})^2$
is the gaussian curvature of the membrane at point $\vec r$.

This clearly displays that the free energy is the sum of two competing terms.
One, proportional to Young's modulus $K_0$, favors configurations where the
membrane buckles out of the plane to screen out the source term.
The other, proportional to the bending rigidity, favors configurations with
minimum mean curvature, that is, the flat state. A buckling instability is
then expected to occur, for a given source term, as the ratio $K_0/\kappa$
increases, or, for given elastic constants, as the strength of the source
increases. Because the stretching energy of isolated dislocations and
disclinations in flat space diverges with system size, buckling occurs
immediately for these defects for all values of the elastic constants in an
infinite system.

The von K\'arm\'an equations
are extremely difficult to solve, and most of what we know on the buckling of
isolated defects was obtained by numerical simulations. One can, however,
make rigorous arguments about the buckling transition, mainly through linear
stability analysis. Observe that, if the energy of a given
configuration $S(\vec r)$ is finite in flat space, then for small but finite
$K_0$ the flat phase will be stable. Then $f\equiv 0$, and one can solve for
the Airy stress function of the flat state $\chi_f$,
$$
\nabla^4 \chi_f = K_0 S(\vec r).
\eqno(10)
$$
Stability with respect to out-of-plane buckling then demands that the linear
eigenvalue problem
$$
\biggl(\kappa \nabla^4  -
{\partial^2 \chi_f \over \partial y^2}{\partial^2  \over \partial x^2}-
{\partial^2 \chi_f \over \partial x^2}{\partial^2  \over \partial y^2}+
2{\partial^2 \chi_f \over \partial x \partial y}
{\partial^2  \over \partial x\partial y} \biggr)f_\eta=\eta f_\eta
\eqno(11)
$$
admit only nonnegative eigenvalues $\eta$.
Equivalently, we see that, as the ratio
$K_0/\kappa$ increases, a buckling instability will
set in as soon as an eigenvalue $\eta$  becomes negative ($\eta=0^-$).
To this eigenvalue corresponds a stable configuration $f_\eta(\vec r)\ne 0$:
the up-down symmetry is now broken. Thus, $f_\eta (\vec r)$ can be regarded as
the order parameter of a Landau theory that describes the buckling transition.
The Landau free energy, given by Eq.~(9), contains terms quadratic and
quartic in $f$. Suppose for simplicity that $K_0$
is fixed (it can always be scaled out of the problem). Then the buckling
instability is reached by decreasing $\kappa$ from infinity to a critical
value, $\kappa^*$.

It is easy to see that the behavior of $f (\vec r)$ near
the critical point is that expected from Landau theory.
Let $\hat f_\eta (\vec r)$ be the normalized eigenfunction of Eq.~(11)
obtained at the critical point $\kappa=\kappa^*$. Below the buckling
transition, but sufficiently close to it ($\kappa=\kappa^*-\delta\kappa$,
$\delta\kappa\ll \kappa^*$),
we look for a solution of the von K\'arm\'an equations of the form
$$
\eqalign{
\chi(\vec r)&=\chi_f(\vec r)+\delta\chi(\vec r) \cr
f(\vec r)&=\alpha \hat f_\eta (\vec r), \cr
}
\eqno(12)
$$
where $\alpha$ is a scalar Ising-like order parameter.
We immediately find
$$
\delta\chi(\vec r)=-\alpha^2K_0\nabla^{-4}\hat K(\vec r),
\eqno(13)
$$
where $\hat K(\vec r)$ is the Gaussian curvature of $\hat f_\eta (\vec r)$.
Multiplying Eq.~(11) on the left by $\alpha^{-1} \hat f_\eta (\vec r)$ and
integrating over all space, we find (putting $\Gamma(\vec r)\equiv
K_0\nabla^{-4}\hat K(\vec r)$)
$$
-\delta\kappa\int d^2r \hat f_\eta \nabla^4 \hat f_\eta
+\alpha^2\int d^2r \hat f_\eta \biggl(
{\partial^2 \Gamma \over \partial y^2}{\partial^2  \over \partial x^2}+
{\partial^2 \Gamma \over \partial x^2}{\partial^2  \over \partial y^2}-
2{\partial^2 \Gamma \over \partial x \partial y}
{\partial^2  \over \partial x\partial y} \biggr)\hat f_\eta=-\eta\equiv 0,
\eqno(14)
$$
and, thus, $\alpha\propto(\delta\kappa)^{1/2}$.

\goodbreak
\bigskip
\centerline{\bf III. GRAIN BOUNDARY BUCKLING}
\medskip
Grain boundaries are interfaces joining continuously two crystalline domains,
$A$ and $B$, of different orientation (see Fig.~2). Consider a reference
lattice, e.g., the infinite extension of domain $A$. Domain $B$ can be
obtained by rotating $A$ through an angle $\theta$ about the vertical.
Consider a closed contour in the reference lattice. If the part that lies in B
is rotated as described above, and the contour  encloses a
segment of length $h$ of the grain boundary, the contour will fail to close by
an amount~[6]
$$
b=2h\sin {\theta\over 2}.
\eqno(15)
$$
Thus, dislocations must lie along the grain boundary, with  Burger's vector
density determined by the tilt angle $\theta$. Dislocations need not all have
the same Burger's vector (so long as it has the proper magnitude and
direction on average), but in the simplest case of a symmetric grain
boundary, the boundary is equivalent to an infinite row of identical,
equally spaced dislocations. We begin by studying this case in detail.

Consider a membrane with a density of defects consisting of a row of
dislocations, lying on the $y$--axis, with Burger's vectors $\vec b$:
$$
S(\vec r)= \biggl(b_x{\partial \over \partial_y} -
b_y{\partial \over \partial_x} \biggr)\delta(x)\sum_{n=-\infty}^{\infty}
\delta(y-nh).
\eqno(16)
$$
The Airy stress function in flat space is obtained from Eq.~(10),
$$
\eqalign{
\chi (\vec r)= -b_yK_0{x|x|\over 4h} +
b_yK_0{1\over 4\pi}&\sum_{n=1}^{\infty}
{1\over n}\exp(-2\pi n|x|/h)x\cos(2\pi ny/h)\cr
-b_xK_0{1\over 4\pi}&\sum_{n=1}^{\infty}
{1\over n^2}\exp(-2\pi n|x|/h)\biggl({h \over 2\pi}+n|x|\biggr)\sin(2\pi ny/h).
}
\eqno(17)
$$

The requirement that the energy per unit length of the boundary be finite
implies, in flat space, the well known condition that there be no net
component of the Burger's vector parallel to the boundary; i.e., $b_y=0$.
The stresses then decay exponentially in the grains as $|x|\to \infty$,
$$
\eqalign{
\sigma_{xx}&\simeq{K_0b\over 2h}\exp\biggl(-{2\pi\over h}|x|\biggr)
\biggl(1+{2\pi\over h}|x|\biggr)\sin\biggl({2\pi\over h}y\biggr)\cr
\sigma_{yy}&\simeq{K_0b\over 2h}\exp\biggl(-{2\pi\over h}|x|\biggr)
\biggl(1-{2\pi\over h}|x|\biggr)\sin\biggl({2\pi\over h}y\biggr)\cr
\sigma_{xy}&\simeq{K_0b\over 2h}\exp\biggl(-{2\pi\over h}|x|\biggr)
x\cos\biggl({2\pi\over h}y\biggr).\cr
}
\eqno(18)
$$
This expression for the stress tensor can be used to perform a stability
analysis according to Eq.~(11). The buckling problem becomes simpler, and more
physically intuitive, if we work in a zero-range approximation, which should be
particularly appropriate for high-angle grain boundaries, where $h$ is of
order a few lattice constants. We expect that this approximation describes the
asymptotic physics more generally in the limit $x\gg h$.
In this approximation, which preserves the integrated stress,
the components of the stress tensor become
$$
\eqalign{
\sigma_{xx}&\approx {K_0b\over\pi}\delta(x)\sin\biggl({2\pi\over h}y\biggr)\cr
\sigma_{yy}&\approx\sigma_{xy}\approx 0.\cr
}
\eqno(19)
$$
Figure~3 illustrates the physical origin of the stresses above, for the case
of a $21.8^o$ symmetric grain boundary in a triangular lattice.

Using Eqs.~(6) and (19), we see that the
stability problem now requires finding nontrivial solutions of the equation
$$
\nabla^4  f- {K_0 b \over \kappa\pi}\delta(x)\sin({2\pi\over h}y)
{\partial^2 f\over \partial x^2}=0.
\eqno(20)
$$
We look for a periodic solution in the direction of the boundary, with period
$h$. Note that $f$ must be of the form (up to an additive constant and
a linear term in $x$)
$$
f(\vec r)={2\pi\over h}\sum_{n=-\infty}^{\infty}
f(q_n)\exp(iq_ny)\exp(-|q_nx|)(1+|q_nx|),
\eqno(21)
$$
with $q_n=2\pi n/h$, which shows that, upon buckling,
the surface remains asymptotically flat as $|x|\to \infty$.
Substituting this expression for $f$ into Eq.~(20) and taking the
Fourier transform, one obtains the following recursion relation for
$q_n^2f(q_n)\equiv\tilde f(q_n)$:
$$
\tilde f(q_{n+1})+{16\pi^2\kappa\over bhK_0}|q_n|\tilde f(q_n)+
\tilde f(q_{n-1})=0.
\eqno(22)
$$
A buckling instability occurs when the determinant of the tridiagonal matrix
defined by the recurrence first vanishes as the amplitude of the diagonal
terms (i.e., of $\kappa$) is decreased from infinity. This happens for
$$
{16\pi^2\kappa\over bhK_0}=0.8316,
\eqno(23)
$$
which yields the estimate of the critical value of the elastic constants for
the buckling of the zero-range model of symmetric grain boundaries
$$
{K_0bh\over\kappa}\approx 190.
\eqno(24)
$$
For fixed elastic constants, this can be regarded as a condition on the tilt
angle, $\theta\approx b/h$. Just below the flat space Kosterlitz-Thouless
melting temperature we have $K_0 b^2 =16\pi T_{KT}$, so grain boundaries will
buckle for $T\le T_{KT}$ provided $\theta<\theta_c\approx 0.27T_{KT}/\kappa$.

The same conclusion, concerning the shape of the membrane after buckling,
could be arrived at qualitatively from Eq.~(2). Indeed, we note that, if
buckling is to reduce the stretching energy, then the gradient of $f$ ought to
approximately cancel the components of the in-plane displacement [see
Eq.~(3)].
Then $|\partial_i f|$, and, thus, $f$ itself,
should decay exponentially (i.e., as the square root of the in-plane stress).

This finding appears in conflict with evidence of a long range displacement
$f$ reported in Ref.~[5], where a paper model of a grain boundary in a
triangular lattice was constructed. The model is pertinent to the limit
$K_0bh/\kappa\gg 1$ because of the large in-plane shear modulus.
The grain boundary buckles, forms
a hinge-like structure with a definite dihedral angle, and displays
intriguing ripples in one of the two grains. Here, we show that finite size
effects and the slight misalignment of the Burger's vectors with the grain
boundary normal affect the conclusions crucially.
Figure~4 displays an atomistic view of the grain boundary studied in Ref.~[5].
Note that one domain is tilted by the same angle as each of the grains of
Fig.~3, but the other is not. Therefore,
it is no longer possible to create a stress-free configuration with coincident
atomic sites: to bring the grains to match at the boundary in flat space,
one has to compress uniformly (i.e., for all $x$) the untilted grain
by a factor ${\sqrt 7}/3$ in the $y$ direction (or to stretch the tilted grain
correspondingly).
If the grain is allowed out of the plane, then it can
develop ripples in such a way that now the projection of the atomic rows on
the plane is compressed, but the equilibrium interatomic distance in 3D space
is restored. In dislocation language, this grain is characterized by
Burger's vectors with a component parallel to the grain. According to Eq.~(17),
we have now $\sigma_{yy}\approx-b_yK_0x/4h|x|$, which is indeed a constant
compressive load on the untilted grain, and tensile load on the tilted one.
When the in-plane stresses are relieved by formation of ripples in the
compressed grain, stretching energy is traded for bending energy. This is
certainly advantageous in a paper model,
which can be easily bent and hardly stretched;
if the grains are infinite, however, and if $\kappa\ne 0$, the
ripples will themselves cost infinite bending energy per unit boundary length.
This ``infinite'' energy can nevertheless still be less than the ``infinite''
energy of the configuration in flat space, which drives the buckling in the
finite model.

Until now, we have dealt with infinite, periodic arrays of
dislocations.  We emphasize that periodicity played a crucial role in our
arguments that grain boundary buckling does not generate any long range
out-of-plane displacement. This may not be, however, the most physically
relevant situation. In fact,
while grain boundaries cannot terminate inside a crystalline
membrane, three or more of them can radiate outward from a common vertex. The
breakdown of the periodicity of the arrays has extremely interesting
consequences. We illustrate them by explicitly considering the case of three
semi-infinite arrays of dislocations radiating outwards from the origin.

Let us begin by computing the shear stress produced by a
half line of dislocations lying on the negative $y$ axis, equally spaced, and
with Burger's vector pointing in the $x$ direction:
$$
\sigma_{xy}= K_0bx\sum_{n=-\infty}^{-1}{x^2-(y-nh)^2\over
\bigl(x^2+(y-nh)^2\bigr)^2}.
\eqno(25)
$$
With the aid of the Euler-Maclaurin summation formula [11],
$$
\sum_{n=1}^{N-1}F(n)=-{\textstyle{1\over2}}\bigl(F(N)+F(0)\bigr)+
\int_0^NdxF(x)+\int_0^N dx
\biggl(-\sum_{n=1}^\infty{\sin2\pi nx\over n\pi}\biggr){d\over dx}F(x),
\eqno(26)
$$
we find a multipole
expansion for the shear stress field (neglecting exponentially small terms):

$$
\sigma_{xy}={b\over h}x\biggl(-h{y\over x^2+y^2}+
{\textstyle {1\over2}}h^2{y^2-x^2\over(x^2+y^2)^2}-
{\textstyle {1\over6}}h^3{y(y^2-3x^2)\over(x^2+y^2)^3}\biggr)
+O(h^3/r^3).
\eqno(27)
$$

When grain boundaries meet at a vertex, the Burger's vectors $\vec b_i$ and
dislocation  spacings $h_i$ cannot be arbitrary. This is because they
determine the angles, through which the crystal orientation is rotated at
each boundary (see Frank's law). Imposing the constraint that the sum of
the Burger's vector densities radiating from the vertex vanish  is
equivalent to requiring that no net disclinicity be present at the vertex,
i.e., that the sum of the monopole terms of the stress field vanish.
We have an additional degree of freedom at our disposal in the construction
of the vertex, which is a rigid shift of the dislocations along one arm. This
does not affect the monopole moment, but can be used to make
the unphysical dipole terms (corresponding to a Burger's vector
which is not a lattice vector)  vanish.
{\it Higher order multipoles, however, are still present.}
Below some critical value of the elastic constants, the vertex is stable
in the flat phase,  because the
long-range stresses it generates are square integrable, and thus, the
stretching energy is finite. Above that critical value,
a buckling transition will take place
to a phase with long range out-of-plane displacements.
A grain boundary network can thus be regarded as a system of point-like
objects, the vertices where the grains intersect, with long range,
anisotropic interactions. The nature of the buckled phase of such a system
is highly nontrivial, and numerical simulations suggest that
spin-glass-like frustration will be important at low temperature.

\bigskip
\goodbreak
\centerline{\bf IV. NUMERICAL SIMULATIONS}
\medskip
We have carried out numerical simulations of grain boundary buckling,
following the method described in Ref.~[2]. Briefly, we work with a 2D
triangular lattice of atoms, which is allowed to bend. In the ideal,
unstrained crystal, each atom is sixfold coordinated, and each bond has unit
length. To represent quenched defects, note that a positive (negative)
disclination corresponds to a fivefold (sevenfold) coordinated atom, a
dislocation is a disclination dipole, and so on.
The stretching energy is defined as a sum over nearest neighbor atoms,
$$
F_s={\textstyle {\sqrt{3}\over4}}K_0\sum_{<i,j>}(|\vec R_i-\vec R_j|-1)^2,
\eqno(28)
$$
where $\vec R$ is the 3D atomic coordinate.
Each elementary triangle is assigned a unit normal, $\vec n_\alpha$, and the
bending energy is defined as a sum over nearest neighbor triangles,
$$
F_s={\textstyle {2\over\sqrt{3}}}\kappa
\sum_{<\alpha,\beta>}(1-\vec n_\alpha\cdot\vec n_\beta).
\eqno(29)
$$
The total energy is minimized, using a conjugate gradient search [12].

First, we have studied the symmetric, high angle grain boundary, whose
unrelaxed configuration in flat space is shown in Fig.~3. Periodic boundary
conditions are imposed along the grain,
and free boundary conditions in the transverse
direction. Figure~5a shows a relaxed configuration after buckling, obtained
with a ratio of the elastic constants $K_0bh/\kappa\approx 300$. We have
determined numerically that the critical value of the elastic constants for
this symmetric grain boundary is
$$
{K_0bh\over\kappa}\simeq 120.
\eqno(30)
$$
This value is smaller than predicted by the zero range model, and is very
close to that which would be obtained for an isolated dislocation in a
membrane of finite size $h$. Note that the vertical
displacement never exceeds one lattice unit, and that the
asymptotically flat state is reached very quickly as we move away from the
grain boundary. For comparison, Fig.~5b shows the results of the relaxation
of a {\it finite}
segment of the same grain boundary; free boundary conditions are
imposed in all directions. Now one can clearly see a long range structure in
the out-of-plane displacement, due to finite size effects; the long range
stresses, caused by terminating the boundary freely, can be estimated from
Eq.~(27). Figure~5c shows the relaxation and buckling of the asymmetric
boundary shown in Fig.~4, again with periodic boundary conditions along the
boundary and free boundary conditions elsewhere. Although the ripples
found in the paper model of
Ref.~[5] are recovered, the overall structure with periodic boundary
conditions is much flatter.

These simple examples teach us that simulations of extensive grain boundary
networks would be quite challenging. Finite size effects  would make it
difficult
to discern the ``real'' effect, due to the breakdown of periodicity at a grain
boundary vertex, from the spurious effect of freely terminating each of the
boundaries at the edge of the sample. Note, however, that the leading
contribution to the vertex stress tensor is the quadrupole moment. A bona fide
localized defect possessing the same leading multipole stress field is a
tightly bound dislocation pair. This is indeed amenable to computer
simulation.

The buckling of disclinations and dislocations  in a tethered membrane
has been investigated numerically by several authors [2,13].
Both types of defects induce large deflections
in the membrane upon buckling. Figure~6 shows the buckling of a tightly
bound  dislocation pair. One can observe large out-of-plane
displacements.

An interesting question to address concerns the shape of a membrane in the
presence of more than one grain boundary vertex, and the interaction which
arises. An argument can be made that the buckling of two defects in
opposite directions is favored by the bending energy (the stretching
energy is less important in the buckled phase)~[4]. Therefore, one may expect
that point defects have an ``antiferromagnetic'' interaction.
The simplest system where this prediction can be tested is a pair of
impurities consisting of atoms of a bigger size, each of which produces a
spherically symmetric compression of the neighboring lattice in the flat
phase, with stresses that decay as $1/r^2$. For computational purposes, these
defects were embedded near the center of an approximately circular membrane
(radius $R=20$ lattice spacings, and $K_0/\kappa\approx 92$, well above
the buckling transition) and the energy was minimized by constraining the
defects to buckle either on the same side or on opposite sides of the flat
reference configuration. Bonds radiating from the impurity atoms had
equilibrium length 1.5 times the bonds in the matrix.
The energy difference between the two cases is taken
as an estimate of the coupling (the exchange $J$ of an Ising model). We found
that $J$ is positive for separations below three lattice constants, and
negative above that (see Fig.~7). Thus, nearest neighbor
and next-nearest neighbor
impurities have a ferromagnetic coupling, which favors buckling in the
same direction, while defects farther apart  have an antiferromagnetic
interaction, consistent with the argument above. Note the similarity with the
frustrated interactions which arise in spin glasses [8]. In real grain
boundary networks, however, we have the additional possibility of a
competing {\it ferromagnetic} interaction between
nodes connected by a buckled grain boundary.

One final feature may be important in the case of grain
boundary vertices, namely, significant anisotropy of the exchange coupling.
While detailed numerical study of the coupling in this more complicated case
is beyond the scope of the
present work, we present the result of a simulation of
pairs of interacting defects, to show that the basic conclusion, i.e., an
antiferromagnetic interaction between defects, is still valid.
Figure 8 displays the topology of a membrane containing two ``nodes,'' each
modeled as a tightly bound dislocation pair. Each pair can
be viewed as a collapsed (i.e., anisotropic) vacancy in a triangular lattice.
The ratio of the elastic constants is $K_0b^2/\kappa\approx 92,$ well above
the buckling transition of an isolated pair, which we have found to occur for
$K_0b^2/\kappa\approx 26.$
The relaxation was effected subject to the constraint that the 5-fold
coordinated atoms marked by thicker circles in Fig.~8
buckle in the same (Fig.~9a) or in opposite (Fig.~9b) direction. The five-fold
atoms nearest to these heavy circles buckle oppositely to their neighbors as
in Fig.~6.
The total energy is 9\% smaller in Fig.~9b, indicating an
antiferromagnetic interaction between these defects for this separation.

\bigskip
\goodbreak
\centerline{\bf V. DISCUSSION}
\medskip
In conclusion, we hope to have clarified the way in which frustration arises
in grain boundary networks. Particularly important are the {\it nodes}, which
lead to long range strain fields and large out-of-plane displacements upon
buckling. An infinite grain boundary by itself produces relatively small,
exponentially screened deflections below the buckling transition. The stronger
``antiferromagnetic'' interaction between the buckled nodes across a grain
competes with a ferromagnetic nodal interaction which presumably arises when
a buckled grain boundary connects two nearest neighbor nodes.

In studies of spin glasses, one often considers  simplified theoretical
models, abstracted from the complexities of real quantum mechanical spins
interacting via, say, a RKKY interaction [8]. Similar simplifications may be
appropriate for grain boundary networks in membranes. A ``direct'' simulation
of this problem might involve quenching a flat space liquid to produce grains,
tethering neighboring particles to prevent further changes in topology, and
then annealing and relaxing the resulting membrane in three dimensions. The
free boundary conditions natural in this approach would lead to additional
long range strain fields (and further warping of the surface) whenever a
boundary terminates at the membrane edge. Sample sizes much larger than the
grain would be required to accurately model the experiment of Ref.~[3],
although even relatively small systems would be of some interest. One might
also try to simulate grain boundary networks in membranes with a {\it
spherical} topology, similar to the liposomes of Ref.~[3].

In view of the complexity of direct simulations, it may be worth considering
simplified models. One approach would be to deliberately insert vacancies and
interstitials into an otherwise perfect membrane (to simulate nodes), and
ignore the exponentially screened grain boundaries entirely. Such a model
would be similar to the random impurity model discussed in Ref.~[4]. For weak
disorder, the equilibrium flat phase [2] survives at finite temperatures (with
microscopic buckling at small scales), although there are indications of a
spin glass phase when the disorder is strong [14]. Glassy behavior at low
temperatures has
been found in computer simulations [15]. One could abstract even further
and simply replace the buckled defects by Ising degrees of freedom with
effective Hamiltonian
$$
{\cal H}_{eff}=-\sum_{i<j}J(|\vec r_i-\vec r_j |)\sigma_i\sigma_j,
\eqno(31)
$$
where $\sigma_i=\pm 1$, the $\{\vec r_i\}$ are random positions within the
membrane, and $J(r)$ has the form shown in Fig.~7. Simple power counting using
Eq.~(2) suggests that the antiferromagnetic tail of $J(r)$ has the form
$J(r)\approx -\kappa/r^2$. Even the behavior of this simple two-dimensional
Ising spin glass is, to the best of our knowledge, unknown.

\bigskip
\goodbreak
\centerline{\bf ACKNOWLEDGEMENTS}
\medskip
We acknowledge helpful discussions with G. Grest and D. Huse.
This work was supported by the National Science Foundation, through Grant No.
DMR91-15491, and through the Harvard Materials Research Laboratory.

\bigskip\bigskip
\goodbreak
\centerline{\bf References}
\medskip
\frenchspacing

\item{[1]}D. R. Nelson and B. I. Halperin, Phys. Rev. B {\bf 19}, 2457 (1979).
\item{[2]}H. S. Seung and D. R. Nelson, Phys. Rev. A {\bf 38}, 1005 (1988).
\item{[3]}M. Mutz, D. Bensimon, and M. J. Brienne, Phys. Rev. Lett. {\bf 37},
923 (1991).
\item{[4]}D. R. Nelson and L. Radzihovsky, Europhys. Lett. {\bf 16}, 79 (1991);
L. Radzihovsky and D. R. Nelson, Phys. Rev. A {\bf 44}, 3525 (1991); D.
Bensimon, D. Mukamel, and L. Peliti, Europhys. Lett. {\bf 18}, 269 (1992).
\item{[5]}D. R. Nelson and L. Radzihovsky, Phys. Rev. A {\bf 46}, 7474 (1992);
this paper also argues that unscreened {\it disclinations}, quenched in by the
polymerization process, would lead to a wrinkled glass. We exclude this
possibility here, and concentrate on grain boundary networks with no
unscreened disclinicity.
\item{[6]}J. P. Hirth and J. Lothe, {\it Theory of Dislocations} (Wiley, Ney
York, 1992).
\item{[7]}A. P. Young, J. D. Reger, and K. Binder, T. in Appl. Phys. {\bf
71}, 355 (1992).
\item{[8]}K. H. Fischer and J. A. Hertz, {\it Spin Glasses}
(Cambridge University Press, Cambridge, 1991).
\item{[9]}A more extensive discussion can be found in Ref. [2]. The basic
equations of the theory of elastic membranes are the same as in the theory
of thin plates; see, e.g.,  L. D. Landau and E. M. Lifshitz,
{\it Theory of Elasticity} (Pergamon, New York, 1970).
\item{[10]}D. R. Nelson and L. Peliti, J.  Phys. (Paris) {\bf 48}, 1085 (1987).
\item{[11]}K. Knopp, {\it Theorie und Anwendung der Unendlichen Reihen}
(Springer, Berlin, 1964), p. 540.
\item{[12]}W. H. Press, B. P. Flannery, S. A. Teukolsky, and W. T. Vetterling,
{\it Numerical Recipes} (Cambridge University Press, Cambridge, 1986), Ch.~10.
\item{[13]}D. C. Morse and T. C. Lubensky, J. de Physique II {\bf 3}, 531
(1993).
\item{[14]}L. Radzihovsky and P. Le Doussal,  J. de Physique I {\bf 2}, 599
(1992).
\item{[15]}Y. Kantor, Europhys. Lett. {\bf 20}, 337 (1992).
\vfill
\eject

\bigskip
\nonfrenchspacing
{\bf Figure captions}
\bigskip
\noindent {\bf Fig.~1}:
The temperature behavior of the elastic constants of a membrane can cause
finite energy defects, such as vacancies, interstitials, or grain boundaries,
to buckle as the temperature is lowered.
Buckling will occur at higher temperatures for defects
characterized by a larger length scale, according to Eq.~(1). Thus, low angle
grain boundaries will  buckle first upon cooling a polymerized grain boundary
network
\bigskip
\noindent {\bf Fig.~2}:
Closure failure for a small angle ($9.4^o$) grain boundary indicating the
presence of dislocations.
\bigskip
\noindent {\bf Fig.~3}:
Symmetric high angle ($21.8^o$) grain boundary in triangular lattice. The
arrows display the points where the lattice is most strained. The pattern
repeats itself indefinitely in the direction of the grain boundary.
\bigskip
\noindent {\bf Fig.~4}:
Asymmetric high angle grain boundary in triangular lattice (see also
Ref.~[5]). For the grains to match, a uniform compressive load must be
applied to the untilted boundary.
\bigskip
\noindent {\bf Fig.~5}:
{\bf a}: Buckling of a $21.8^o$ symmetric grain boundary (the configuration
in flat space is that of Fig.~3a). Periodic boundary conditions have been
applied in the direction of the grain boundary.
{\bf b}: Same as ({\bf a}), except free boundary conditions have been
applied in all directions.
{\bf c}: Buckling of asymmetric grain boundary (see Fig.~4 for the
configuration in flat space). Periodic boundary conditions have been
applied in the direction of the grain boundary.
\bigskip
\noindent {\bf Fig.~6}:
Buckled tightly bound pair of dislocations with opposite Burger's vectors
(the configuration in flat space is shown in inset).
\bigskip
\noindent {\bf Fig.~7}:
Energy splitting between ``antiferromagnetic'' and ``ferromagnetic''
configurations of two impurities in an otherwise ideal crystalline membrane.
Each impurity is assigned a preferred bond length which is 50\% larger than the
lattice constant. An isolated impurity has a buckling transition for
$K_0/\kappa\ge 8$; here, we have taken $K_0/\kappa=92$ (the lattice spacing is
set to unity). When the impurities are farther apart than two lattice
sites, buckling in opposite direction is energetically favored. The energy
is given in units of the bending rigidity $\kappa$.
\bigskip
\noindent {\bf Fig.~8}:
Two tightly bound dislocation pairs in flat space. Explicitly marked are the
5-fold and 7-fold coordinated atoms. The unevenly spaced atoms reside in
regions of high local strain.
\bigskip
\noindent {\bf Fig.~9a,b}:
Relaxed configuration of a membrane comprising two tightly bound
dislocation pairs (see Fig.~8 for the unrelaxed configuration). The relaxation
was effected subject to the constraint that the 5-fold
coordinated atoms (marked by a thicker circle in Fig.~8)
buckle (a) in the same or (b) in opposite direction.
The total energy is 9\% smaller in the latter case, indicating an
``antiferromagnetic'' interaction between defects.

\vfill
\eject

\bye